\documentclass[journal=jacsat,manuscript=article]{achemso}

\usepackage[version=3]{mhchem} 

\author{Yao-Kung Huang}

\author{Szu-Chao Chen}
\affiliation[Cheng Kung University]
{Department of Physics, National Cheng Kung University, Taiwan}

\author{Yen-Hung Ho}
\affiliation[University of Houston]
{Department of Physics and Texas Center for Superconductivity, University of Houston, Texas, U.S.A.}

\author{Chiun-Yan Lin}
\email{l28981084@mail.ncku.edu.tw}
\phone{+886-6-275-7575}
\fax{+886-6-74-7995}
\affiliation[Cheng Kung University]
{Department of Physics, National Cheng Kung University, Taiwan}

\author{Ming-Fa Lin}
\email{mflin@mail.ncku.edu.tw}
\phone{+886-6-275-7575}
\fax{+886-6-74-7995}
\affiliation[Cheng Kung University]
{Department of Physics, National Cheng Kung University, Taiwan}

\title[An \textsf{achemso} demo]
  {Feature-Rich Magnetic Quantization in Sliding Bilayer Graphenes}

\abbreviations{LL,EF}

\begin{document}

\begin{tocentry}

Some journals require a graphical entry for the Table of Contents.
This should be laid out ``print ready'' so that the sizing of the
text is correct.

Inside the \texttt{tocentry} environment, the font used is Helvetica
8\,pt, as required by \emph{Journal of the American Chemical
Society}.

The surrounding frame is 9\,cm by 3.5\,cm, which is the maximum
permitted for  \emph{Journal of the American Chemical Society}
graphical table of content entries. The box will not resize if the
content is too big: instead it will overflow the edge of the box.

This box and the associated title will always be printed on a
separate page at the end of the document.

\end{tocentry}

\begin{abstract}
  The generalized tight-binding model, based on the subenvelope functions of distinct sublattices, is developed to investigate the magnetic quantization in sliding bilayer graphenes. The relative shift of two graphene layers induces a dramatic transformation between the Dirac-cone structure and the parabolic band structure, and thus leads to drastic changes of Landau levels (LLs) in the spatial symmetry, initial formation energy, intergroup anti-crossing, state degeneracy and semiconductor-metal transition. There exist three kinds of LLs, i.e., well-behaved, perturbed and undefined LLs, which are characterized by a specific mode, a main mode plus side modes, and a disordered mode, respectively. Such LLs are clearly revealed in diverse magneto-optical selection rules. Specially, the undefined LLs frequently exhibit intergroup anti-crossings in the field-dependent energy spectra, and show a large number of absorption peaks without optical selection rules.

\end{abstract}

\section{Introduction}

The essential electronic properties of layered graphenes can be easily modulated by means of the stacking configurations,$^{1-9}$ external fields,$^{9,10}$ numbers of layers,$^{1,2}$ deformed structures,$^{11-13}$ and so on. The highly symmetric layer configurations include the, AA$^{14}$ AB$^{15,16}$ and ABC stackings.$^{17,18}$ The random-stacking $^{19-21}$ and the twisted$^{22-24}$ structures can be produced by specific experimental methods. These systems possess very special band structures, which can be further quantized by applying a uniform perpendicular magnetic field $\bf{B}$ = B$_{0}\hat{z}$.$^{25-30}$ The diverse magneto-electronic energy spectra are clearly displayed in the interesting physical properties, e.g., the optical selection rules$^{31-33}$ and the novel Hall effect.$^{34-37}$ This work is focused on how the various stacking configurations in sliding bilayer graphenes lead to the feature-rich magnetic quantization and the complex magneto-absorption spectra.

The quantized Landau levels (LLs) are strongly dependent on the stacking configuration. A multi-layer graphene exhibits N groups of valence and conduction LLs.$^{26}$ Each group in the AA-stacked system has the well-behaved LLs with monolayer-like energy spectra and spatially symmetric wave functions.$^{31}$ All the LLs of the AB-stacked system can be regarded as alike to the  bilayer and monolayer ones.$^{26,27}$ Only a few LL intergroup anticrossings appear in the $B_{0}$-dependent energy spectra when the field strength is sufficiently large.$^{27}$ Both intergroup and intragroup anticrossings are revealed in the ABC-stacked systems; furthermore, the latter arise from the Sombrero-shaped energy bands induced by the interlayer couplings.$^{38}$ The majority of the LLs in these two systems are well-behaved ones, and the anticrossing LLs within a certain $B_{0}$-range belong to the perturbed ones. By changing the chiral angle, fractal LL energy spectra exist in twisted bilayer graphenes.$^{30}$

The effective-mass model may be too complex or cumbersome for solving the equations governing the magnetic-electronic properties of sliding bilayer graphenes with random stacking configurations, although it could conceivably be used to comprehend the low-energy magnetic quantization of fewer-layer graphenes.$^{27,39}$ Apparently, the dramatic transformation between the Dirac cone structure and the parabolic band structure$^{5,6}$ means that the magnetic Hamiltonian matrix will be too complicated to diagonalize. We have developed the generalized tight-binding model for various external fields,$^{9,25,26,31,32,38}$ in which the Hamiltonian matrix is built from the tight-binding basis functions, i.e., the subenvelope functions on the distinct sublattices. The main characteristics of the geometric structures are directly reflected in the magnetically quantized electronic states. The spatial distributions of the subenvelope function are critical in distinguishing how many kinds of LLs exist in sliding bilayer graphenes. This study shows that the well-behaved, the perturbed and the undefined LLs are  characterized by a specific mode, a main mode plus side modes, and a disordered mode, respectively. The first theoretical predictions on the undefined LLs indicate that many absorption peaks without specific selection rules are included.

\section{Method}
The stacking configurations can be tuned by the relative shift between two graphene layers. The AA stacking, corresponding to $\delta=0$ (in units of the C$-$C bond length $b_{0}=1.42$ $\AA$), has identical (x,y) projections for all carbon atoms.
When one layer is gradually shifted along the armchair direction ($\hat{x}$), the AB stacking configuration is reached when $\delta=1$. With a further increase in the shift, $\delta=1.5$ is defined as the AA$^{\prime}$ stacking, where each carbon atom has the same chemical environment, but a different coordinate projection. The low-energy Hamiltonian, associated with four 2p$_{z}$ orbitals in the primitive unit cell, can be expressed as H=
$-\sum\limits_{<i,j>}$ $\gamma_ {ij}$
$c^{\dag}_{i}$  $c_{j}$. $\gamma_ {ij}$ is the intralayer or the interlayer hopping integral between lattice sites i and j;  its strength presumably depends on the distance and the angle of two 2$p_{z}$ orbitals in the form of$^{6}$
\begin{equation}
-\gamma_{ij}=
\gamma_{0}
e^{-\frac{d-b_{0}}{\varrho}}
[1-(\frac{\textbf{d}\cdot\textbf{e}_{z}}{d})^{2}]
+\gamma_{1}
e^{-\frac{d-d_{0}}{\varrho}}
(\frac{\textbf{d}\cdot \textbf{e}_{z}}{d})^{2}
\end{equation}
$\gamma_{0}=-2.7$ eV is the intralayer nearest-neighbor hopping integral, $\gamma_{1}=0.48$ eV the interlayer vertical hopping integral, $\textbf{d}$ the position vector connecting two lattice points, $d_{0}=3.35$ $\AA$ the interlayer distance, and $\varrho=0.184 b_{0}$ the decay length.

The magnetic field $B_{0}\hat{z}$ induces periodical Peierls phases which modulate the hopping integral as $\gamma_{ij}$ ($\textbf{B})=\gamma_{ij}$
exp$(i\frac{2\pi}{\Phi_{0}}
\int_{r_{i}}^{r_{j}}
\textbf{A}(\textbf{r})\cdot d\textbf{r})$,$^{25,26,47}$ where  $\Phi_{0}$ (= hc/e) is the flux quantum. Within a chosen vector potential $\textbf{A}= (0,B_{0}x,0)$, the primitive unit cell is an enlarged rectangle with 8$R=8\times79000/B_{0}$ carbon atoms, where $R$ is the ratio of the flux quantum to the magnetic flux through each hexagon, and restricted to a positive integer in the calculations. The magnetic Bloch wave function is the linear superposition of the $8R$ tight-binding functions
$\{$$\lvert A_{ik}^{l}\rangle$; $\lvert B_{ik}^{l}\rangle$ $\arrowvert$ $^{l=1,2}_{i=1,2,..,2R}\}$(Fig. 1(a)), where $l$ is the layer index, and $i$ represents the $i$th atom in each sublattice.

\section{Results and discussion}
\subsection{Electronic properties}
The low-lying band structures in the sliding bilayer graphenes intricately respond to the various stacking configurations. They are equivalent for two valleys K$^{+}$ and K$^{-}$, since the space-inversion symmetry is unbroken.$^{40}$
The AA-stacked system, as shown in Fig. 2(a) for energy bands near K$^{+}$, possesses two pairs of linear conduction and valence bands, in which the first and the second pairs of Dirac cones are situated at E$^{c,v}\sim$ +0.32 eV and $-$0.36 eV, respectively.
As the configuration gradually moves away from an AA stacking configuration (Fig. 2(b) at $\delta=1/8$), the electronic states in the lower cone of the first pair strongly hybridize with those of the upper cone of the second pair. An eye-shaped stateless region along the $k_x$-direction is created near E$_{F}$, and two-band contact points remain as the Fermi-momentum states (k$_{F}$'s).$^{6}$ Electronic states near these two k$_{F}$'s are symmetric about the k$_{y}$-axis, mainly owing to the space-inversion and y$\longrightarrow$$-$y symmetries. It should be noted that the Dirac cone structures are becoming smoother and distorted during the variation of the stacking configuration, and are even separated for a $\delta$ larger than the critical displacement ($\delta_c\simeq5/8$; $\delta$ = 6/8 in Fig. 2(c)). The complete separations of the upper and the lower Dirac  cones indicate that two pairs of energy bands are reformed at the different energies. The band-edge states, corresponding to  the first and the second pairs of energy bands, have $E^{c,v}\simeq0$ and  ($E^c\simeq0.32$ eV; $E^v\simeq-0.36$ eV), respectively.

A further increase from $\delta=1$ to $\delta=1.5$ also leads to drastic changes of the band structure. As is apparent from Fig. 2(d), the AB stacking exhibits two pairs of parabolic bands, being characterized by the weak  band overlap near $E_F$. With increments of $\delta$, the parabolic bands of the first pair are seriously distorted along $\hat k_y$ and $-\hat k_y$ simultaneously, as shown in Fig. 2(e) at $\delta=11/8$. The region outside the created eye-shape region, with two Dirac points at distinct energies, grows quickly; furthermore, two neighboring conduction (valence) bands form strong hybrids. Finally, the two pairs of the isotropic Dirac cones are reformed in the AA$^\prime$ stacking (Fig. 2(f)), in which the Dirac points are located at different wave vectors with $E^{c,v}\simeq-0.11$ eV and 0.1 eV. The cone axes are tilted in the opposite directions for  the conduction and valence bands, as clearly displayed in the two distinct loops with a constant energy measured from the current Dirac point (Fig. 1(c)).

Each LL wave function is characterized by the subenvelope functions of distinct sublattices. The AA stacking exhibits well-behaved LLs with four-fold degeneracy for each $(k_{x},k_{y})$ state. All the LLs localized at the 2/6 position of the enlarged unit cell are shown in Fig. 3(a) for $B_{0}=40$ T; similar localization centers corresponding to the other degenerate states occur at the 1/6, 4/6 and 5/6 positions. The localization centers are determined by the effective  momentum due to the magnetic field and the $k_{y}-$component of the $K^{+}$ or $K^{-}$ point (Fig. 1(b)). Two groups of conduction and valence LLs are defined by the number (n) of zero points in the dominating $B^{1}$ or $B^{2}$ sublattice. The subenvelope functions are well fitted by the composite functions of the Hermite polynominal and the Gaussian function; they can also be obtained from the low-energy expansion around the $K^{+}$ point by diagonalizing the Hamiltonian matrix. The $n_{1}^{c,v}=0$ LL of the first group (blue) starts to form at $E^{c,v}\simeq0.32$ eV, and then the LL energies grow in the $\sqrt {n_{1}^{c,v}}$ form.
Similar results can also be found in the second group (red) with the $n_{2}^{c,v}=0$ LL at $E^{c,v}\simeq -0.36$ eV.

The feature-rich LLs are generated by the relative shift of two graphene layers. The low-lying LLs near $E_{F}$ are drastically altered even for a small shift, e.g., the eight LLs are equally contributed by the first and the second groups in Fig. 3(b) at $\delta=1/8$. The proportional relationship between their energies and $\sqrt{n^{c,v}}$ is thoroughly destroyed, mainly owing to the strong hybridization of the two neighboring Dirac-cone states (Fig. 2(b)). The subenvelope functions, corresponding to the eight LLs in the range of $-0.1$ eV $\leq$$E^{c,v}$$\leq$ 0.1 eV, originate from the magnetic quantization of the electronic states outside the eye-shaped stateless region. They still have a specific zero-point number, as indicated by the $B^{1}$ or $B^{2}$ sublattice. However, the regular oscillations with spatial symmetries are somewhat distorted. Such perturbed LLs can be described by a main  mode ($n^{c,v}$) and extra modes ($n^{c,v}=\pm1$; even $\pm2$ in a few cases). The relationship between these two kinds of mixed modes is mainly determined by the field strength and the stacking configuration. The extra modes can be easily enhanced by modulating $B_{0}$, which thus leads to the low-energy intergroup LL anticrosssings (Fig. 4(b)).

The main characteristics of the two groups of LLs are totally changed when the relative shift is sufficiently large. The simple relationship between LL energy and $\sqrt{n^{c,v}}$ is absent for all LLs, as shown in Fig. 3(c) at  $\delta=6/8$. The first and second groups start to form at $E_{1}^{c,v}(n_{1})\simeq 0$ and $(E^{c}(n_{2})\simeq 0.32$ eV $\&$ $E^{v}(n_{2})\simeq -0.36$ eV), respectively, and directly reflect the band-edge state energies in Fig. 2(c). Only two LLs in the first group are close to $E_{F}$, and the energy spacings of the other LLs exceed 0.1 eV. This means that only these two LLs can occupy all electronic states outside the eye-shaped region. The conduction and valence LLs of the first group belong to the perturbed LLs before the emergence of those of the second group. For example, the two LLs near $E_{F}$ are roughly equivalent to $n_{1}^{v}=0$  and $n_{1}^{c}=1$ from the dominating $B^{1}$  sublattice. On the other hand, all the LLs in the second group, as well as the neighboring LLs in the first group, oscillate irregularly on distinct sublattices and cannot be identified as any simple mode. The spatial symmetries of the subenvelope functions are lost; furthermore, the irregular oscillations continuously change with the field strength (Fig. 5(b)). It is quite difficult to characterize such LLs based on the spatial distributions of the dominating sublattice; therefore, the highly-degenerate electronic states without main modes are classified as undefined LLs. Several of the initial LLs in the second group exhibit rather strong fluctuations, but not the regular oscillations with few zero points. This clearly illustrates that the undefined LLs are closely related to the thorough separations of the upper and the lower Dirac cones (Fig. 2(c)). Such special LLs are easily observed in the 5/8 $\leq\delta \leq$ 7/8 region corresponding to the drastic deformation of the Dirac-cone structures.

Two pairs of parabolic bands in the AB stacking are quantized into two groups of well-behaved LLs, as shown in Fig. 3(d). The first group possesses $n_{1}^{v}$ = 0 and $n_{1}^{c}$ = 1 LLs near $E_{F}$ as a result of the weak band overlap. Both groups of LLs can exist simultaneously at higher energy, while their spatial symmetries are totally different from each other. Consequently, there are no intergroup LL anti-crossings in the $B_{0}$-dependent spectrum (Fig. 4(d)). Concerning the $\delta=11/8$ bilayer graphene, the strong hybridization of the two neighboring conduction (valence) bands results in many undefined LLs at $|E^{c,v}|\geq 0.25$ eV (Fig. 3(e)). They can survive in the  range of 10.5/8 $\leq\delta\leq$ 11.5/8, reflecting the dramatic transformation from the parabolic bands into the Dirac cones. Some perturbed LLs with few zero points appear at lower energy, which means that the initial LLs in both groups are generated from the quantized states of the deformed Dirac cones. An obvious energy gap $E_{g}$$\simeq$0.2 eV, related to the energy difference of two Dirac points, is revealed. When the Dirac cone structures are reconstructed, the AA$^{\prime}$ stacking possesses two groups of well-defined LLs at $E^{c,v}(n_{1}=0)$ = $-$0.11 eV and $E^{c,v}(n_{2}=0)$ = 0.1 eV. The proportional relationship between LL energy and $\sqrt{n^{c,v}}$ is recovered. The two localization centers of the $n_{1}^{c,v}$ = 0 and $n_{2}^{c,v}$ = 0 LLs deviate from the 2/6 position along opposite directions since the $k_{y}$-component is different for the current Dirac points and the $K^{+}$ point. Corresponding to the centered LL, the tilted Dirac-cone axis causes the conduction and the valence LLs in each group to have the opposite deviations, and this tendency is  gradually enhanced with an increase in energy.

The $B_{0}$-dependent energy spectra are useful in understanding the formation of the two groups, the intergroup anticrossings, the energy gap and the state degeneracy. From the initial LLs at $B_{0}\to 0$, the first and second groups in the AA system are formed at 0.32 eV and $-$0.36 eV, respectively (Fig. 4(a)). All the LLs, with the exception of the fixed $n_{1}^{c,v} = 0$ and $n_{2}^{c,v} = 0$ LLs, own a simple $\sqrt{B_{0}}$-dependence. The energy gap, i.e., the energy difference between the two LLs nearest to $E_{F}$, exhibits frequent semiconductor-metal transitions. State degeneracy remains four-fold, except at the intergroup crossings. The small shift shown in Fig. 4(b) at $\delta$ = 1/8 induces the special intergroup anticrossings from the low-lying perturbed LLs in $-$0.1 eV $\leq E^{c,v}\leq 0.1$ eV. Therefore, there exist two entangled LLs very close to $E_{F}$, which dominate the semiconductor-metal transitions. Their zero points gradually decrease with an increasing $B_{0}$. These transitions are associated with the distorted band structure outside but near the eye-shaped region (Fig. 2(b)). Moreover, the space-inversion and y$\longrightarrow$$-$y symmetries near two $k_{F}$'s lead to eight-fold degenerate LLs in the $-$0.05 eV $\leq E^{c,v}\leq 0.05$ eV region for $B_{0}\leq 10$ T.

A dramatic transformation between the Dirac-cone structure and the parabolic band structure appears at larger shifts, and so do the first and second group. Both groups, shown in Fig. 4(c) at $\delta =6/8$, are reformed at $E^{c,v}(n_{1})\simeq 0$ and ($E^{c}(n_{2}) = 0.32$ eV $\&$ $E^{v}(n_{2})=-0.36$ eV); similar LL distributions can also be found in the 5/8 $\leq\delta\leq 8/8$ region. Each LL in the second group displays significant anticrossings with all the LLs in the first group, which clearly illustrates that the former is composed of various zero-point modes, or that it can be verified as an undefined LL. The energy gap is non-existent for $B_{0}\leq 30$ T and gradually appears and increases for $B_{0}\geq 30$ T. As for the eight-fold degenerate LLs, the range of existence is extended into $-$0.1 eV $\leq E^{c,v}\leq  0.1$ eV and $B_{0}\leq 10$ T. However, this range narrows for a further increase of $\delta$ and disappears for $\delta\ge1$.

During the stacking configuration transformation from AB $\to$ AA$^{\prime}$, the conduction (valence) LLs of the first group and the valence (conduction) LLs of the second group approach each other, as indicated in Figs. 4(d)-4(f). The LLs conglomerate and are reformed as the first (second) group until at least $\delta\geq 10.5/8$. The reformation of both groups causes the semiconductor-metal transition to occur more frequently in the $B_{0}$-dependent energy spectra (Fig. 4(e)), while the AB stacking only exhibits a relatively slow gap opening. There are also many intergroup anticrossings from the undefined LLs at $|E^{c,v}| \geq 0.25$ eV, but only a few from the perturbed LLs at lower energy (green circles). Apparently, such anticrossings are absent in the AB, AA$^{\prime}$ and AA stacking configurations.

It is noteworthy how the spatial distributions of the LL wave functions are dramatically altered in the intergroup anticrossings. The low-lying anticrossings of the perturbed LLs for $\delta=1/8$ are illustrated in Fig. 5(a). The $n_{2}^{c} = 3$ LL of the second group (red circles) avoids crossings with the $n_{1}^{v} = 6$ and 5 LLs (blue circles) in the range of 26 T $\le\,B_0\le\, 50$ T. At $B_{0}=26$ T, the main $n_{2}^{c} = 3$ mode only somewhat deviates from the regular distribution on the dominating sublattice, or the side modes of $n_{2}^{c} = 4$ and $n_{2}^{c} = 5$ are weak. With an increase in field strength, the deviation becomes more pronounced and eventually reaches a maximum at $B_{0}\simeq 31.5$ T, where the anticrossing $n_{1}^{v} = 6$ LL displays a similar spatial variation. Their subenvelop functions possess the identical side modes of $n = 4$ and $n = 5$; therefore, they are forbidden to have the same energy. A more obvious anticrossing, related to the $n_{2}^{c} = 3$ and $n_{1}^{v} = 5$ LLs, occurs at $B_{0}\simeq 34$ T; this is mainly caused by the $n = 4$ side mode. Moreover, the anticrossing of $n_{2}^{c} = 3$ and $n_{1}^{v} = 7$ is quite faint at $B_{0}\simeq 28$ T, further indicating that the $n$ $\pm$ 1 side modes are stronger than the $n$ $\pm 2$ ones. As a result, the $n_{2}^{c} = 3$ LL does not show anticrossings with the $n_{1}^{v}$ $\geq 8$ LLs. In addition, the few low-lying anticrossings at $\delta = 11/8$ mainly originate from  the side  modes of n $\pm$ 1 (Fig. 4(e)).

On the other side, the undefined LLs for $\delta = 6/8$ and $\delta = 11/8$, respectively shown in Figs. 5(b)
and 5(c) exhibit relatively strong oscillations without spatial symmetry for any field strength. The number of oscillations is large even for the three initial undefined LLs of the second group ($\alpha,\beta;\gamma$), and it gradually grows as $B_{0}$ decreases. That is to say, they do not have a main mode in the entire $B_{0}$-range. However, the main modes of the perturbed LLs can survive even in the intergroup anticrossings.

\subsection{Optical properties}

The diverse spatial distributions of three kinds of LLs are directly reflected in the magneto-optical absorption spectra calculated from the Fermi golden rule.$^{31,32,41}$ The spectral intensities are dominated by the intralayer nearest-neighbor hopping integral; that is, they are associated with the $A^l (B^l)$ sublattice of the initial state and the $B^l (A^l)$ sublattice of the final state. The AA stacking, as shown in Fig. 6(a) at $B_{0}=40$ T, displays many prominent symmetric peaks that obey the specific selection rule $\Delta n=n_{m}^{v}-n_{m}^{c}= \pm1$. This rule can be understood from the well-behaved spatial distributions of the subenvelope functions on the intralayer A$^{l}$ and B$^{l}$ sublattices.$^{31,32}$ There are only two categories of absorption peaks arising from intragroup optical vertical excitations. Intergroup excitations are forbidden because of the special relationships among the wave functions of the two Dirac cone structures.$^{41}$ Two threshold peaks, associated with the four LLs nearest to $E_{F}$, occur at $\omega=0.05$ eV, and the other intragroup excitation peaks come into existence at $\omega \geq$ 0.63 eV. Consequently, a wide zone of forbidden frequencies is observed.

\subsection{Conclusion}
We further developed the generalized tight-binding model based on the subenvelope functions of distinct sublattices. This model is very suitable for understanding the magnetic quantization of sliding bilayer graphenes, and could also be used to investigate the magnetic properties in the other 2D materials, such as MoS$_{2}$$^{42,43}$ and Silicene.$^{44,45}$ The relative shift of two graphene layers induces a transformation between two types of band structures; therefore, feature-rich LLs are revealed in the field-dependent energy spectra by the initial formation energies, intergroup anti-crossings, state degeneracy, semiconductor-metal transitions and localization centers. The spatial distribution symmetries of three kinds of LLs dominate the diverse magneto-optical selection rules. More peculiarly, the undefined LLs display the anti-crossing energy spectra for any field strength, and create many absorption peaks in the absence of a specific selection rule. The predicted magneto-electronic properties and absorption spectra could be verified by scanning tunneling spectroscopy$^{7,46}$ and optical spectroscopy,$^{33}$ respectively.


\newpage
\renewcommand{\baselinestretch}{0.2}

 \begin{itemize}
 
 \item[1. ] Castro, Neto, A. H. et al. The electronic properties of graphene. $Rev.$ $Mod.$ $Phys.$ \textbf{81}, 109-162 (2009).
 \item[2. ] Gruneis, A. et al. Tight-binding description of the quasiparticle dispersion of graphite and few-layer graphene $Phys.$ $Rev.$ $B.$ \textbf{78}, 205425 (2008).   
 \item[3. ] Mak, K. F., Shan, J., Heinz, T. F. Atomically ThinMoS$_{2}$: A New Direct-Gap Semiconductor. $Phys.$ $Rev.$ $Let.t$ \textbf{104}, 176404 (2010).
 \item[4. ] Lobato, I., Partoens, B. Multiple Dirac particles in AA-stacked graphite and multilayers of graphene. $Phys.$ $Rev.$ $B$ \textbf{83}, 165429 (2011).
 \item[5. ] Son, Y. W. et al. Electronic topological transition in sliding bilayer graphene $Phys.$ $Rev.$ $B$ \textbf{84}, 155410 (2011).
 \item[6. ] Koshino, M. Electronic transmission through AB-BA domain boundary in bilayer graphene. $Phys.$ $Rev.$ $B$ \textbf{88}, 115409 (2013).
 \item[7. ] Li, G. H. et al. Observation of Van Hove singularities in twisted graphene layers. $Nature$ \textbf{6}, 109-113 (2010).
 \item[8. ] Kim, K. S. et al. Coexisting massive and massless Dirac fermions in symmetry-broken bilayer graphene. $Nat.$ $Materials$ \textbf{12}, 887-892 (2013). 
 \item[9. ] Lu, C. L. et al. Influence of an electric field on the optical properties of few-layer graphene with AB stacking. $Phys.$ $Rev.$ $B$ \textbf{73}, 144427 (2006).
 \item[10.] Taisuke, O., Aaron, B., Thomas, S. Controlling the electronic structure of bilayer graphene. $Science$ \textbf{313}, 951-954 (2006).
 \item[11.] Lim, C. H. Y. X. et al. A hydrothermal anvil made of graphene nanobubbles on diamond. $Nat.$ $Communications$ \textbf{4}, 1556 (2013).
 \item[12.] Bao, W. et al. Controlled ripple texturing of suspended graphene and ultrathin graphite membranes. $Nat.$ $Nanotechnol.$ \textbf{4} 562-566 (2009).
 \item[13.] Rainis, D. et al. Gauge fields and interferometry in folded graphene. $Phys.$ $Rev.$ $B$ \textbf{83}, 165403 (2011).
 \item[14.] Lee, J. K. et al. The growth of AA graphite on (111) diamond. J $Chem.$ $Phys.$ \textbf{129}, 234709 (2008).
 \item[15.] Reina, A. et al. Large area, few-layer graphene films on arbitrary substrates by chemical vapor deposition. $Nano$ $Lett.$ \textbf{9}, 30-35 (2009).
 \item[16.] Ferrari, A. C. et al. Raman spectrum of graphene and graphene layers. $Phys.$ $Rev.$ $Lett.$ \textbf{97}, 187401 (2006).
 \item[17.] Lui, C. H. et al. Imaging stacking order in few-layer graphene. $Nano$ $Lett$ \textbf{11}, 164-169 (2011).
 \item[18.] Norimatsu, W., Kusunoki, M. Selective formation of ABC-stacked graphene layers on SiC(0001). $Phys.$ $Rev.$ $B$ \textbf{81}, 161410(R) (2010).
 \item[19.] Zhang, W. et al. Molecular adsorption induces the transformation of rhombohedral- to Bernal-stacking order in trilayer graphene. $Nat.$ $Communications$ \textbf{4}, 2074-2081 (2013).
 \item[20.] San-Jose, P. et al. Stacking Boundaries and Transport in Bilayer Graphene $Nano$ $Lett.$ \textbf{14}, 2052-2057 (2014).
 \item[21.] Hattendorf, S., Georgi, A., Liebmann, M., Morgenstern, M. Networks of ABA and ABC stacked graphene on mica observed by scanning tunneling microscopy. $Surf.$ $Sci.$ \textbf{610}, 53-58 (2013). 
 \item[22.] Iwasaki, T. et al. Formation and structural analysis of twisted bilayer graphene on Ni(111) thin flims. $Surf.$ $Sci.$ \textbf{625}, 44-49 (2014)
 \item[23.] Kim, S. M. et al. Synthesis of Patched or Stacked Graphene and hBN Flakes: A Route to Hybrid Structure Discovery. $Nano$ $Lett.$ \textbf{13}, 933-941 (2013).
 \item[24.] Brown, L. et al. Twinning and Twisting of Tri- and Bilayer Graphene $Nano$ $Lett.$ \textbf{12}, 1609-1615 (2012).
 \item[25.] Ho, J. H. et al. Magnetoelectronic Properties of a Single-Layer Graphite $J.$ $Phys.$ $Soc.$ $Jpn.$ \textbf{75}, 114703 (2006). 
 \item[26.] Lai, Y. H., Ho, J. H., Chang, C. P., Lin, M. F. Magnetoelectronic properties of bilayer Bernal graphene. $Phys.$ $Rev.$ $B$ \textbf{77}, 085426 (2008).
 \item[27.] Koshino, M., McCann, E. Landau level spectra and the quantum Hall effect of multilayer graphene $Phys.$ $Rev.$ $B$ \textbf{83}, 165443 (2011).
 \item[28.] McCann, E., Fal$^{,}$ko, V. I. Landau-Level Degeneracy and Quantum Hall Effect in a Graphite Bilayer. $Phys.$ $Rev.$ $Lett.$ \textbf{96}, 086805 (2006).
 \item[29.] Trellakis, A. Nonperturbative Solution for Bloch Electrons in Constant Magnetic Fields. $Phys.$ $Rev.$ $Lett.$ \textbf{91}, 056405 (2003).
 \item[30.] Wang, Z. F., Liu, F., Chou, M. Y. Fractal Landau-Level Spectra in Twisted Bilayer Graphene. $Nano$ $Lett.$ \textbf{12}, 3833-3839 (2012).
 \item[31.] Ho, Y. H. et al. Optical transitions between Landau levels: AA -stacked bilayer graphene. $Appl.$ $Phys.$ $Lett.$ \textbf{97}, 101905-101907 (2010).
 \item[32.] Ho, Y. H. et al. Magneto-optical Selection Rules in
 Bilayer Bernal Graphene. $ACS$ $Nano$ \textbf{4}, 1465-1472 (2010).
 \item[33.] Henriksen, E. A. et al. Cyclotron Resonance in Bilayer Graphene. $Phys.$ $Rev.$ $Lett.$ \textbf{100}, 087403 (2008).
 \item[34.] Zhang, Y. B., Tan, Y. W., Stormer, H. L., Kim, P. Experimental observation of the quantum Hall effect and Berry's phase in graphene. $Nature$ \textbf{438}, 201-204 (2005).
 \item[35.] Novoselov, K. S. et al. Unconventional quantum Hall effect and Berry's phase of 2£k in bilayer graphene. $Nat.$ $Phys.$ \textbf{2}, 177-180 (2006).
 \item[36.] Taychatanapat, T., Watanabe, K., Taniguchi, T., Jarillo-Herrero, P. Quantum Hall effect and Landau-level crossing of Dirac fermions in trilayer graphene. $Nat.$ $Phys.$ \textbf{7}, 621-625 (2008).
 \item[37.] Zhang, L. et al. The experimental observation of quantum Hall effect of l=3 chiral quasiparticles in trilayer graphene. $Nat.$ $Phys.$ \textbf{7}, 953-957 (2011).
 \item[38.] Lin, C. Y. et al. Stacking-dependent magneto-electronic properties in multilayer graphenes. Summited to Carbon.
 \item[39.] Sena, S. H. R., Pereira, J. M. Jr., Peeters, F. M., Farias, G. A. Landau levels in asymmetric graphene trilayers. $Phys.$ $Rev.$ $B$ \textbf{84}, 205448 (2011).
 \item[40.] Koshino, M., McCann, E. Parity and valley degeneracy in multilayer graphene $Phys.$ $Rev.$ $B$ \textbf{81}, 115315 (2010).
 \item[41.] Chiu, C. W. et al. Critical optical properties of AA-stacked multilayer graphenes $Appl.$ $Phys.$ $Lett.$ \textbf{103}, 041907-041911 (2013). 
 \item[42.] Mak, K. F., Lee, C., Hone, J., Shan, J., Heinz, T. F. Atomically ThinMoS2: A New Direct-Gap Semiconductor. $Phys.$ $Rev.$ $Lett.
 $ \textbf{105}, 136805 (2010).
 \item[43.] Wang, Q. H. et al. Electronics and optoelectronics of two-dimensional transition metal dichalcogenides. $Nat.$ $Nanotechnol.$ \textbf{7}, 699-712 (2012).
 \item[44.] Feng, B. J. et al. Evidence of Silicene in Honeycomb Structures of Silicon on Ag(111). $Nano$ $Lett.$ \textbf{12}, 3507-3511 (2012).
 \item[45.] Vogt, P. et al. Silicene: Compelling Experimental Evidence for Graphenelike Two-Dimensional Silicon. $Phys.$ $Rev.$ $Lett.$ \textbf{108}, 155501 (2012).
 \item[46.] Song, Y. J. et al. High-resolution tunnelling spectroscopy of a graphene quartet $Nature$ \textbf{467}, 185-189 (2010).
 \item[47.] Luttinger, J. M. The Effect of a Magnetic Field on Electrons in a Periodic Potential. $Phys.$ $Rev.$ \textbf{84}, 814 (1951).
 
 \end{itemize}

\begin{flushleft}
\textbf{Acknowledgement}
\hspace{1mm}
This work was supported in part by the National Science Council of Taiwan,
the Republic of China, under Grant Nos. NSC 98-2112-M-006-013-MY4 and NSC 99-2112-M-165-001-MY3.
\end{flushleft}
\begin{flushleft}
\textbf{Author contributions}
\hspace{1mm}
M.F.L. conceived this research.
Y.K.H. performed the major part of the calculations.
S.C.C. and Y.H.H. performed the theoretical derivations.
Y.K.H., S.C.C., Y.H.H. and C.Y.L. calculated the optical absorption spectra.
All authors discussed the results. 
Y.K.H., C.Y.L. and M.F.L. composed the manuscript.
\end{flushleft}
\begin{flushleft}
\textbf{Competing Interests}
\hspace{1mm}
The authors declare that they have no competing financial interests.
\end{flushleft}
\begin{flushleft}
\textbf{Correspondence}
\hspace{1mm}
Correspondence and requests for materials should be addressed to C.Y.L. (email: l28981084@mail.ncku.edu.tw)
and M.F.L. (email: mflin@mail.ncku.edu.tw).
\end{flushleft}

\newpage
\begin{flushleft}
\textbf{Figure 1:} (a) The sliding bilayer graphene with a relative shift along the armchair direction has an enlarged rectangular unit cell in a uniform magnetic field. (b) The first Brillouin zone with two corners $K^{+}$ and $K^{-}$. (c) The constant energy curves of $\delta\,=12/8$ correspond to the current Dirac points.
\end{flushleft}

\begin{flushleft}
\textbf{Figure 2:} The low-lying band structures near the $K^{+}$ point for various stacking configurations: (a) $\delta\,=0$, (b) 1/8, (c) 6/8, (d) 8/8, (e) 11/8; (f) 12/8.
\end{flushleft}

\begin{flushleft}
\textbf{Figure 3:} The LL energies and wave functions at $B_{0}=40$ T for (a) $\delta\,=0$, (b) 1/8, (c) 6/8, (d) 8/8, (e) 11/8; (f) 12/8.
\end{flushleft}

\begin{flushleft}
\textbf{Figure 4:} The magnetic-field-dependent energy spectra for
(a) $\delta\,=0$, (b) 1/8, (c) 6/8, (d) 8/8, (e) 11/8; (f) 12/8.
\end{flushleft}

\begin{flushleft}
\textbf{Figure 5:} The evolution of the subenvelope functions on the dominating $B^{1}$ sublattice during the intergroup anticrossing for (a) $\delta\,=1/8$, (b) 6/8; (c) 11/8. $\alpha$, $\beta$ and $\gamma$ represent the initial three undefined LLs of the second group.
\end{flushleft}

\begin{flushleft}
\textbf{Figure 6:} The magneto-absorption spectra at $B_{0}=40$ T for (a) $\delta\,=0$, (b) 1/8, (c) 6/8, (d) 8/8, (e) 11/8; (f) 12/8.
\end{flushleft}

\begin{figure}
\includegraphics[scale=0.45]{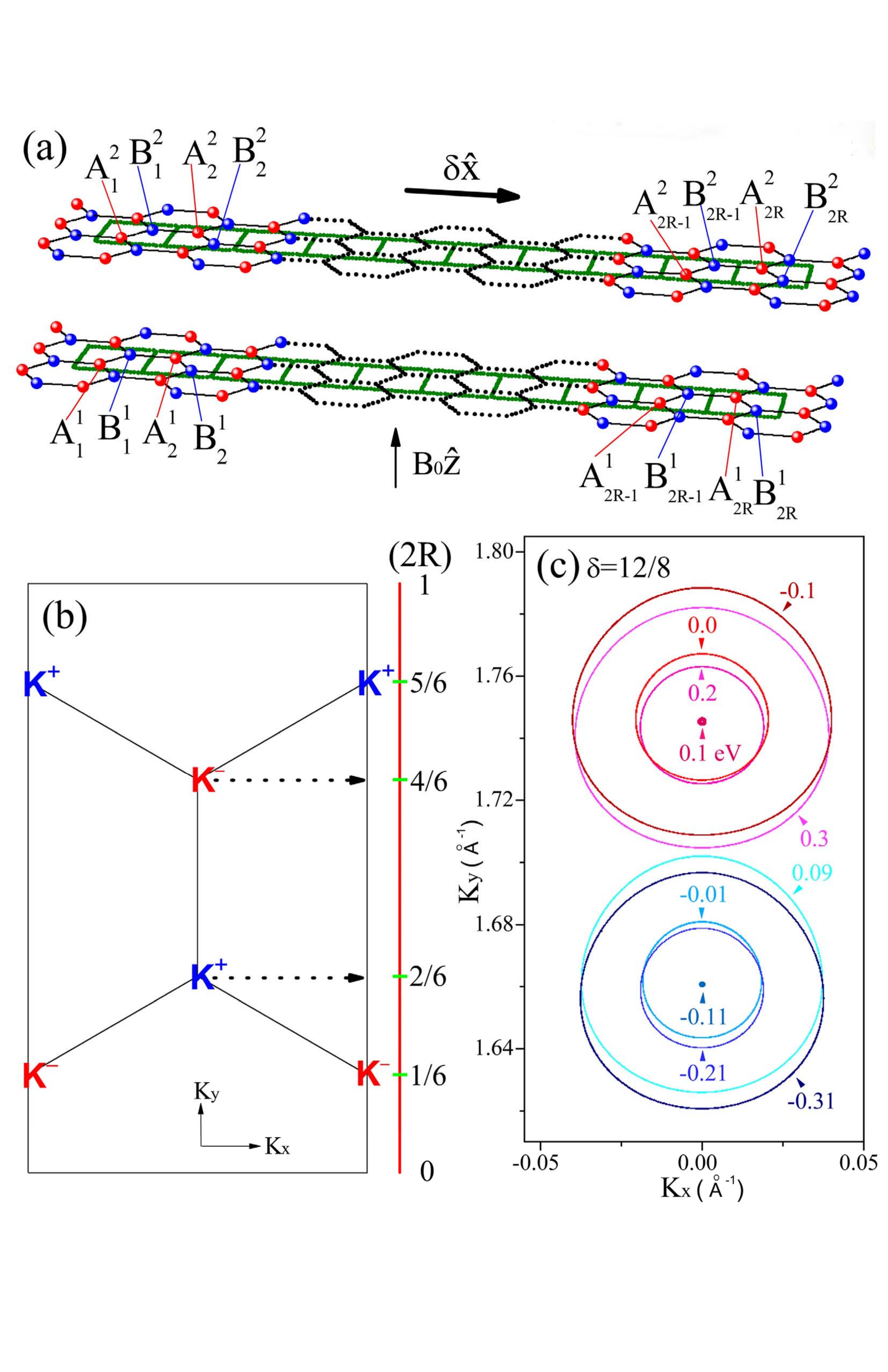}
\caption{ (a) The sliding bilayer graphene with a relative shift along the armchair direction has an enlarged rectangular unit cell in a uniform magnetic field. (b) The first Brillouin zone with two corners $K^{+}$ and $K^{-}$. (c) The constant energy curves of $\delta\,=12/8$ correspond to the current Dirac points.}
\end{figure}

\begin{figure}
\includegraphics[scale=0.45]{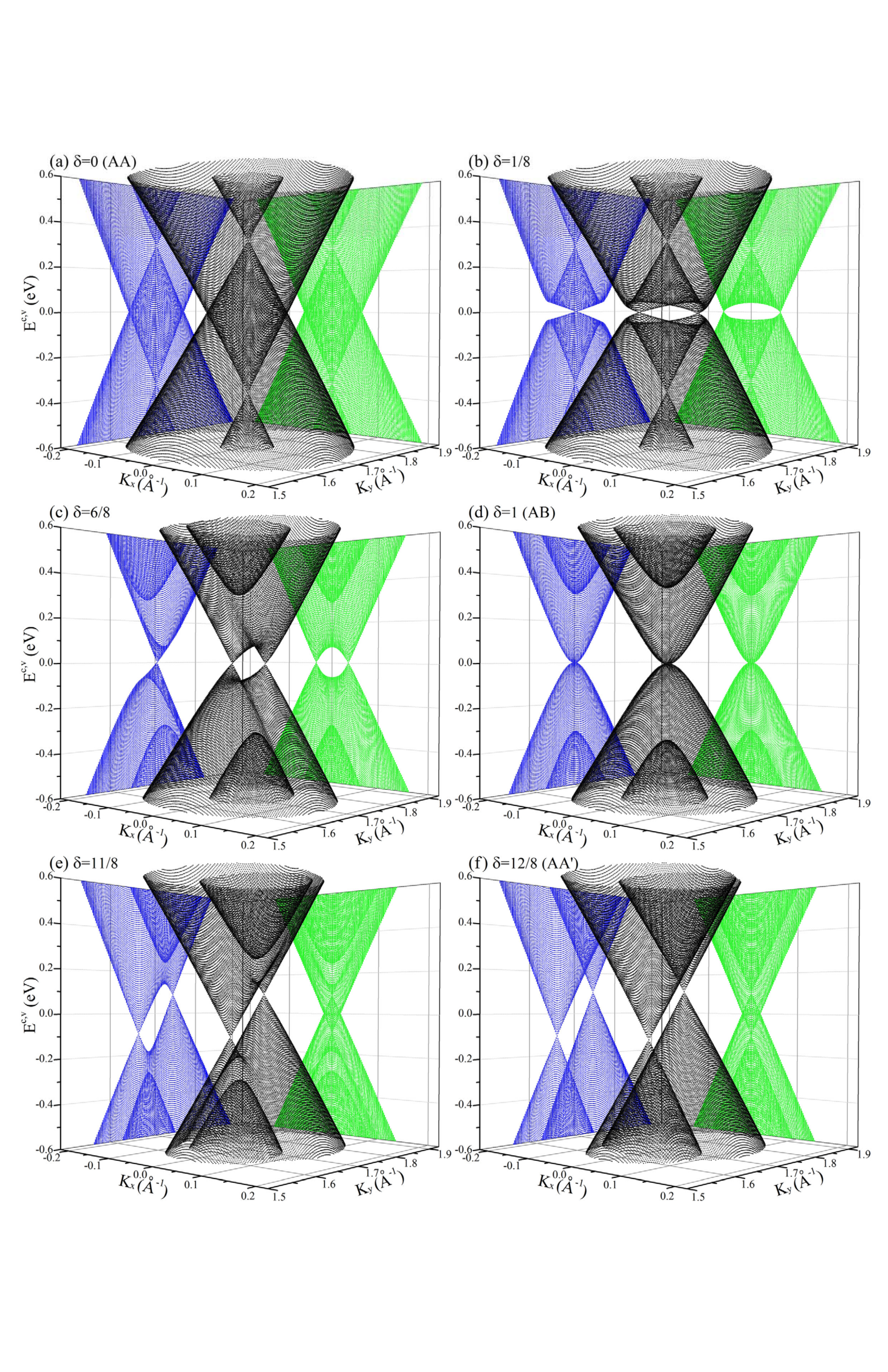}
\caption{ The low-lying band structures near the $K^{+}$ point for various stacking configurations: (a) $\delta\,=0$, (b) 1/8, (c) 6/8, (d) 8/8, (e) 11/8; (f) 12/8.}
\end{figure}

\begin{figure}
\includegraphics[scale=0.45]{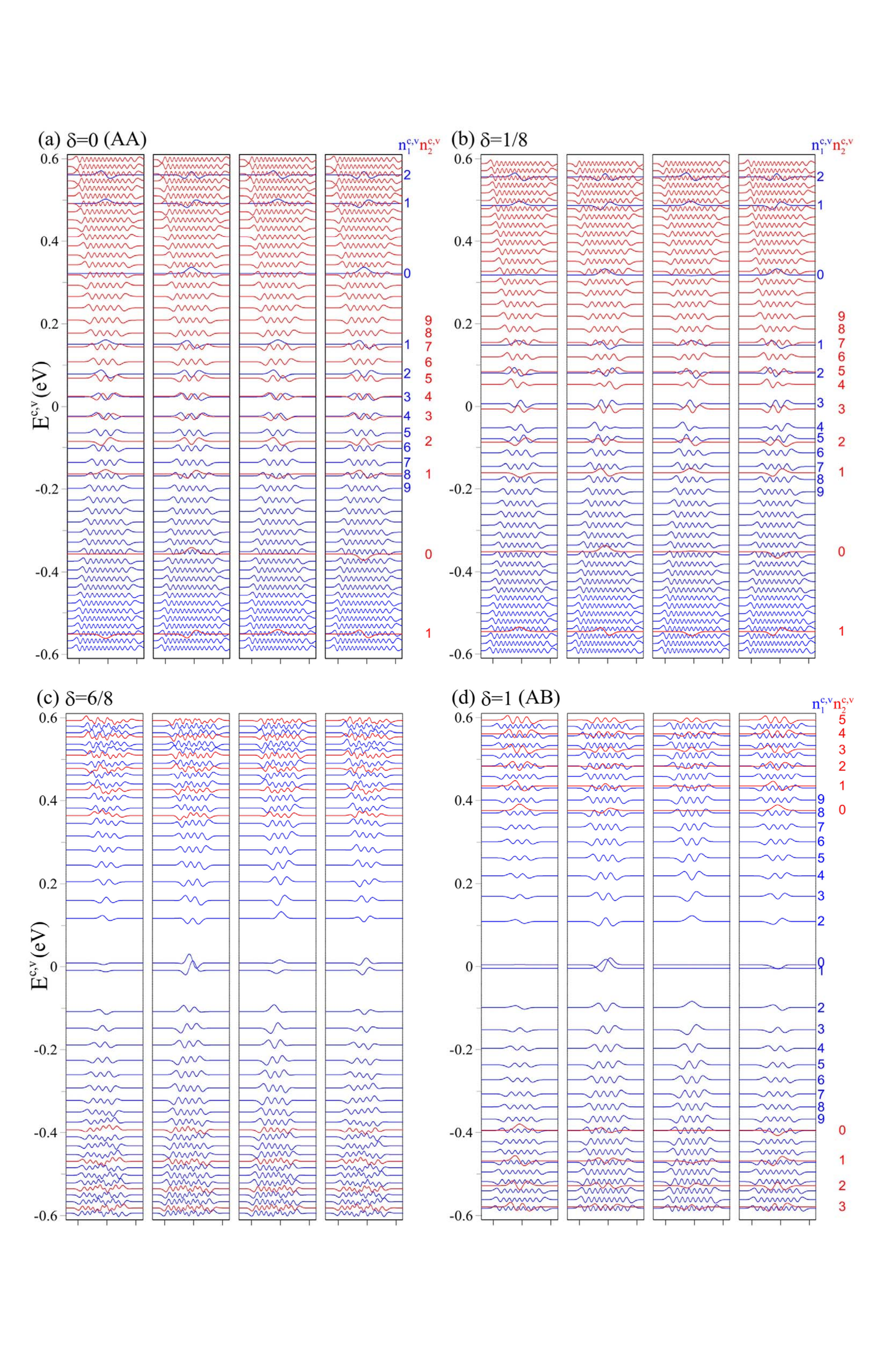}

\end{figure}

\begin{figure}
\includegraphics[scale=0.45]{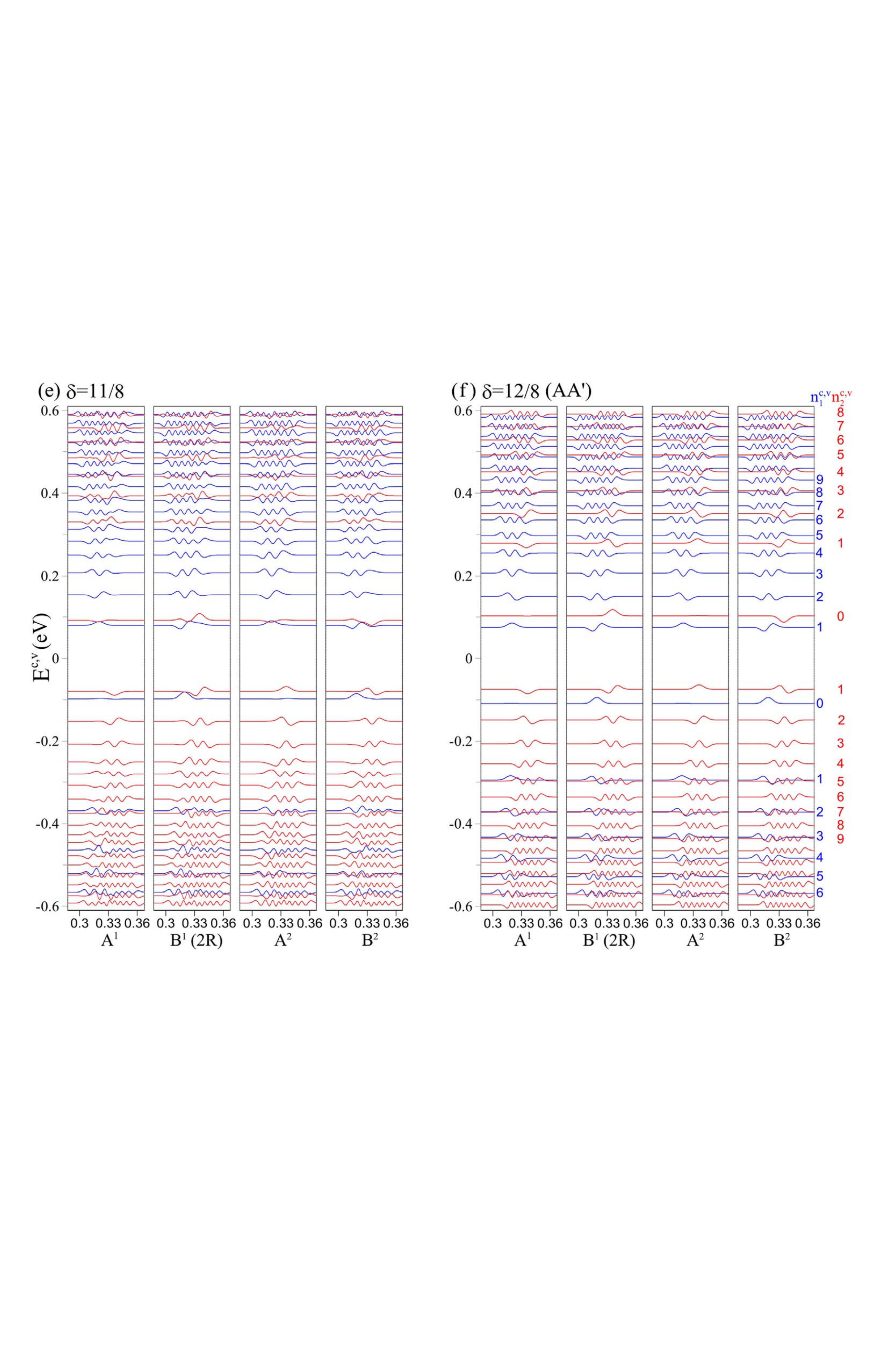} \caption{The LL energies and wave functions at $B_{0}=40$ T for (a) $\delta\,=0$, (b) 1/8, (c) 6/8, (d) 8/8, (e) 11/8; (f) 12/8.}
\end{figure}

\begin{figure}
\includegraphics[scale=0.45]{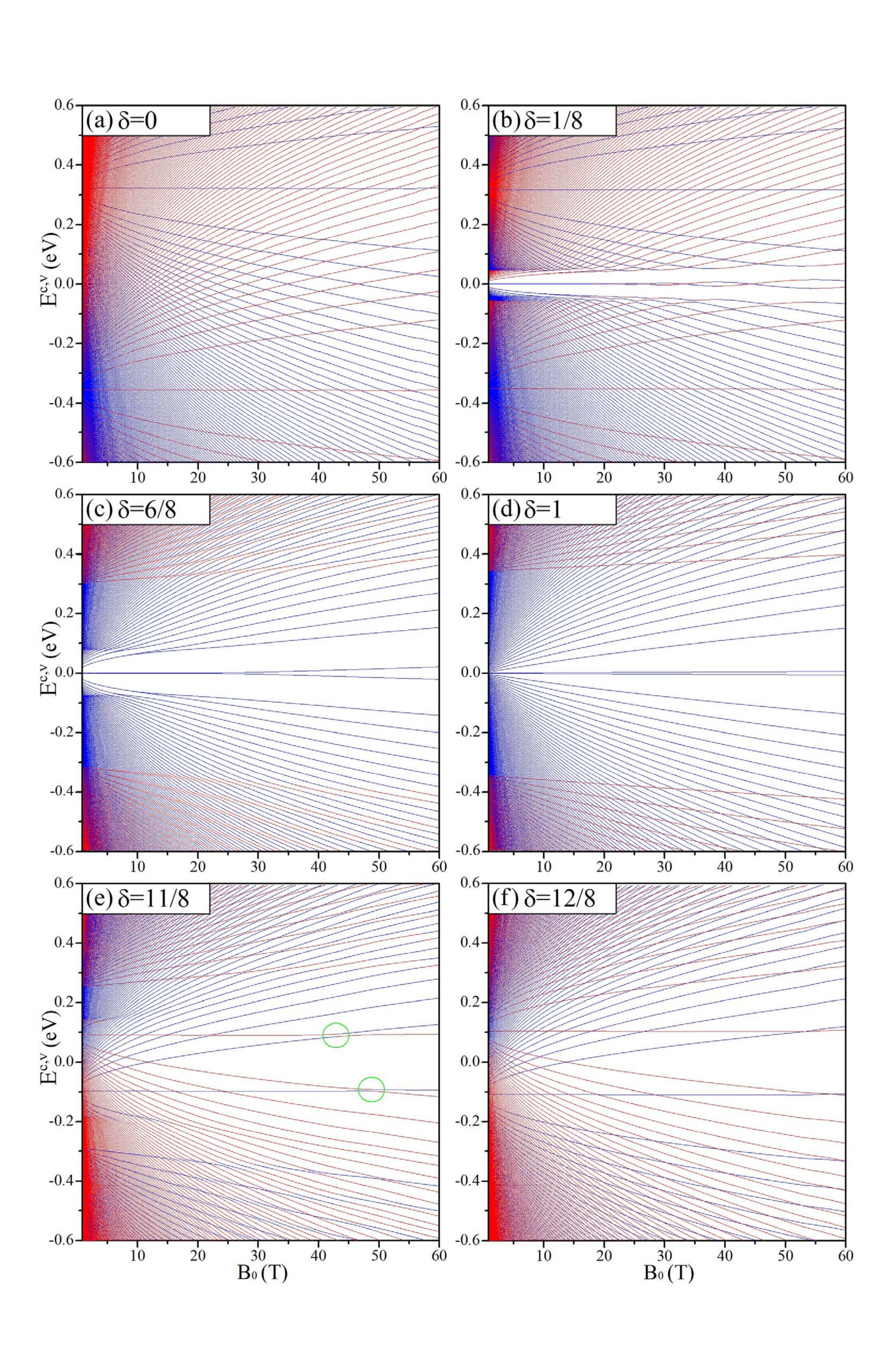} \caption{The magnetic-field-dependent energy spectra for
(a) $\delta\,=0$, (b) 1/8, (c) 6/8, (d) 8/8, (e) 11/8; (f) 12/8.}
\end{figure}

\begin{figure}
\includegraphics[scale=0.45]{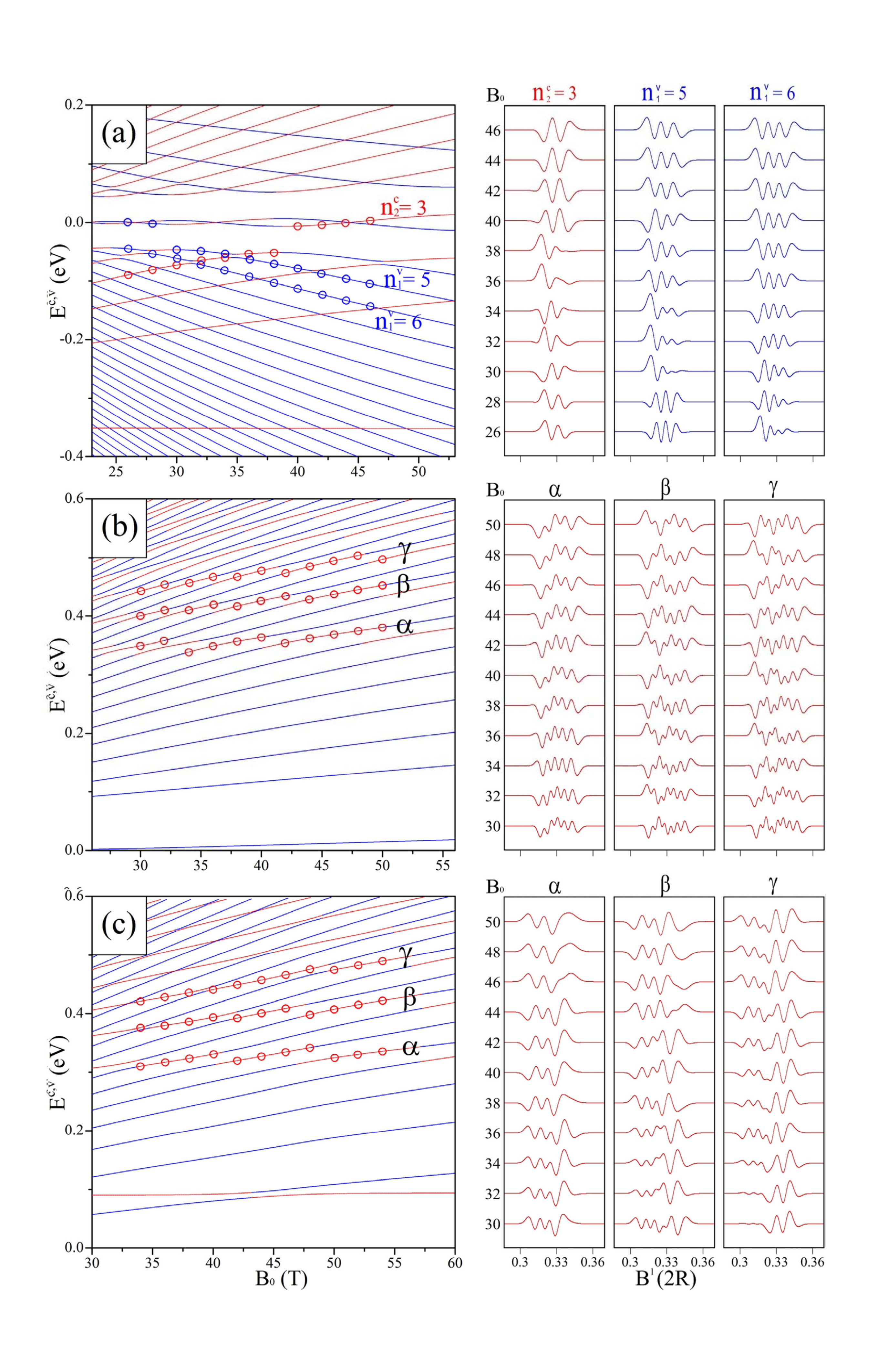} \caption{The evolution of the subenvelope functions on the dominating $B^{1}$ sublattice during the intergroup anticrossing for (a) $\delta\,=1/8$, (b) 6/8; (c) 11/8. $\alpha$, $\beta$ and $\gamma$ represent the initial three undefined LLs of the second group.}
\end{figure}

\begin{figure}
\includegraphics[scale=0.45]{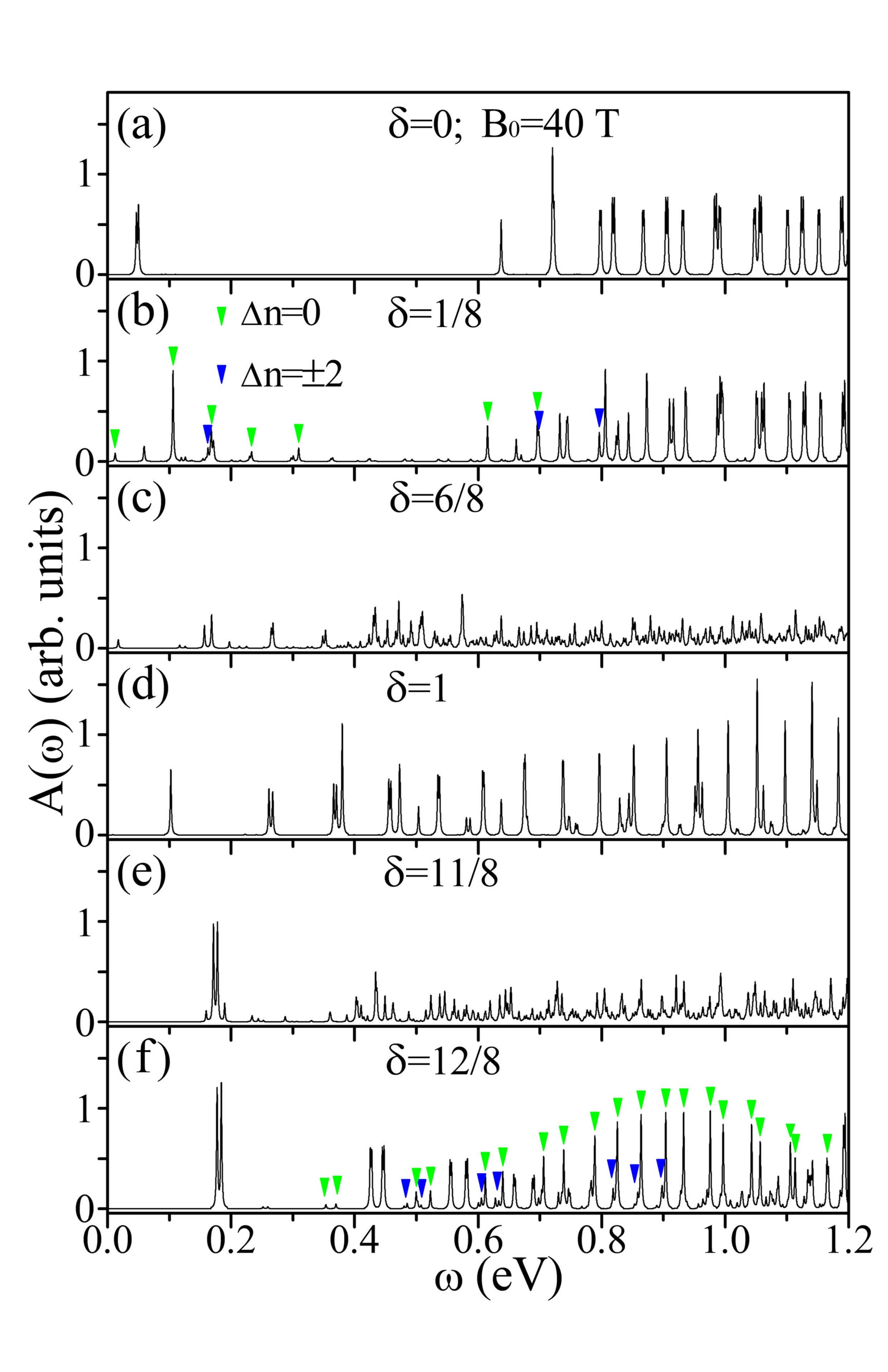} \caption{The magneto-absorption spectra at $B_{0}=40$ T for (a) $\delta\,=0$, (b) 1/8, (c) 6/8, (d) 8/8, (e) 11/8; (f) 12/8.}
\end{figure}





\bibliography{achemso-demo}

\end{document}